\documentclass[12pt]{article}

\usepackage[utf8]{inputenc}
\usepackage{graphicx}
\usepackage{dcolumn}
\usepackage{bm}
\usepackage{subcaption} 
\usepackage{amsmath}
\usepackage{amssymb}
\usepackage{eurosym}
\usepackage{tcolorbox}
\usepackage[normalem]{ulem}
\usepackage{xcolor}

\usepackage[margin=1.2in]{geometry} 

\newcommand\LH[1]{#1}
\newcommand\removedLH[1]{}

\title{Beyond the Aggregated Paradigm:\\ Phenology and Structure in Mutualistic Networks}

\author{
    Clàudia Payrató-Borràs$^{a,b}$,
    Carlos Gracia-Lázaro$^{b}$,\\
    Laura Hernández$^{a}$, and
    Yamir Moreno$^{b,c,d}$
}

\date{}

\begin{document}

\maketitle

\begin{center}
  \begin{small}
    \noindent $^a$ Laboratoire de Physique Théorique et Modélisation, UMR CNRS, Université de Cergy-Pontoise, 2 Avenue Adolphe Chauvin, F-95302, Cergy-Pontoise Cedex, France\\
    $^b$ Institute for Biocomputation and Physics of Complex Systems (BIFI),\\University of Zaragoza, Spain\\
    $^c$ Department of Theoretical Physics, Faculty of Sciences, University of Zaragoza, Spain\\
    $^d$ Centai Institute, Turin, Italy\\
  \end{small}
\end{center}

\vspace{5mm}

\begin{abstract}
\begin{tcolorbox}
Mutualistic interactions, where species interact to obtain mutual benefits, constitute an essential component of natural ecosystems. The use of ecological networks to represent the species and their ecological interactions allows the study of structural and dynamic patterns common to different ecosystems. However, by neglecting the temporal dimension of mutualistic communities, relevant insights into the organization and functioning of natural ecosystems can be lost. Therefore, it is crucial to incorporate empirical phenology -the cycles of species' activity within a season- to fully understand the effects of temporal variability on network architecture. In this paper, by using two empirical datasets together with a set of synthetic models, we propose a framework to characterize phenology on ecological networks and assess the effect of temporal variability. Analyses reveal that non-trivial information is missed when portraying the network of interactions as static, which leads to overestimating the value of fundamental structural features. We discuss the implications of our findings for mutualistic relationships and intra-guild competition for common resources. We show that recorded interactions and species' activity duration are pivotal factors in accurately replicating observed patterns within mutualistic communities. Furthermore, our exploration of synthetic models underscores the system-specific character of the mechanisms driving phenology, increasing our understanding of the complexities of natural ecosystems.
\end{tcolorbox}
\end{abstract}

\bigskip


%
%
%
%
%

\section{Introduction}
Mutualistic relationships, which for centuries had been mainly overlooked as fascinating but marginally relevant, are today known to play a crucial role in shaping natural ecosystems\cite{bascompte2013mutualistic}. In mutualistic communities, individuals from different species interact through activities that\removedLH{report} \LH{provide} them mutual benefits such as food, shelter or an increase in reproductive success --being pollination of plants by animals the most characteristic example. The explosion of network science in the 90s fostered the development of a holistic understanding of these communities using the language of ecological networks\cite{heleno2014ecological}, which allowed investigating the existence of widespread structural or dynamical patterns despite ecosystem's differences in species composition, climate or location\cite{jordano1985ciclo,bascompte2003nested,thebault2010stability}. 

In the context of community ecology, the network's nodes typically represent different species, while the links (which might be weighted or binary) depict the ecological interactions among them. In the particular case of mutualistic networks, the nodes may be moreover separated into two different sets or guilds --e.g. plants and pollinators-- \LH{such} that\removedLH{in the simplest case only held} \LH{the links representing the mutualistic interaction only exist between nodes of different guilds}\removedLH{links among them}, resulting in a so-called bipartite network. This representation can then be enhanced, for instance, by adding other kinds of ecological interactions like competition for common resources, which then leads to a bilayered depiction of the community\cite{gracia2018joint}. 

Here, we will attempt to understand how by neglecting the temporal dimension of real mutualistic communities, that is, by working with the aggregated version of the interaction networks, we can lose relevant insight into the organization and functioning of natural ecosystems. In this sense, we will incorporate into the network formalism the information about empirical phenology, namely the biological activity cycles of species that, to a certain extent, constrain and articulate how ecological relationships occur among individuals. This endeavor, \LH{although largely overlooked,} is\removedLH{certainly} not entirely new\cite{olesen2008temporal,encinas2012phenology,sajjad2017effect,ramos2018phenology,chacoff2018interaction} and, in fact, during recent years the need of moving towards a more realistic depiction of ecological communities has been stressed more than once\cite{ings2009ecological, heleno2014ecological}. All in all, aggregated networks are still commonly used in the characterization of the structure and dynamics of mutualistic systems, \LH{probably due to the scarcity of data covering different years with information both about the observed interactions and the phenology}. Along this paper we will address this problem by examining the effects on the perceived network architecture of taking into account -at least partly- the temporal dimension of mutualism, using information extracted from two empirical datasets provided by\cite{burkle2013plant} and \cite{kantsa2018disentangling}, \LH{which do include simultaneously the dynamical information mentioned above}. Moreover, we will not only consider their consequences for mutualistic relationships but also for intra-guild competition for common resources.

Accordingly, the present paper is organized as follows: first, we complete this introduction by reviewing some basic notions and recent advances in the study of phenology; next, we propose a framework to characterize empirical phenology\removedLH{as well as} \LH{and we propose} a set of synthetic models\removedLH{that permit assessing and contrasting} \LH{in order to study} the effect of temporal variability beyond the scarce number of open datasets; third, we present\removedLH{ the results of structurally characterizing} \LH{a structural characterization of }  two real datasets\removedLH{and how their networks of interactions vary along} \LH{taking into account the variation of their interaction networks during}  the season and compare these observations with the explorations produced by\removedLH{some} synthetic 
models \LH{that include stylized characteristics of the phenology}; and finally, we conclude with a discussion of these results and some unresolved, future challenges.

\subsection{From cherry blossom to digital cameras}

\begin{figure}[h!]
\begin{center}
\includegraphics[width=0.8\textwidth]{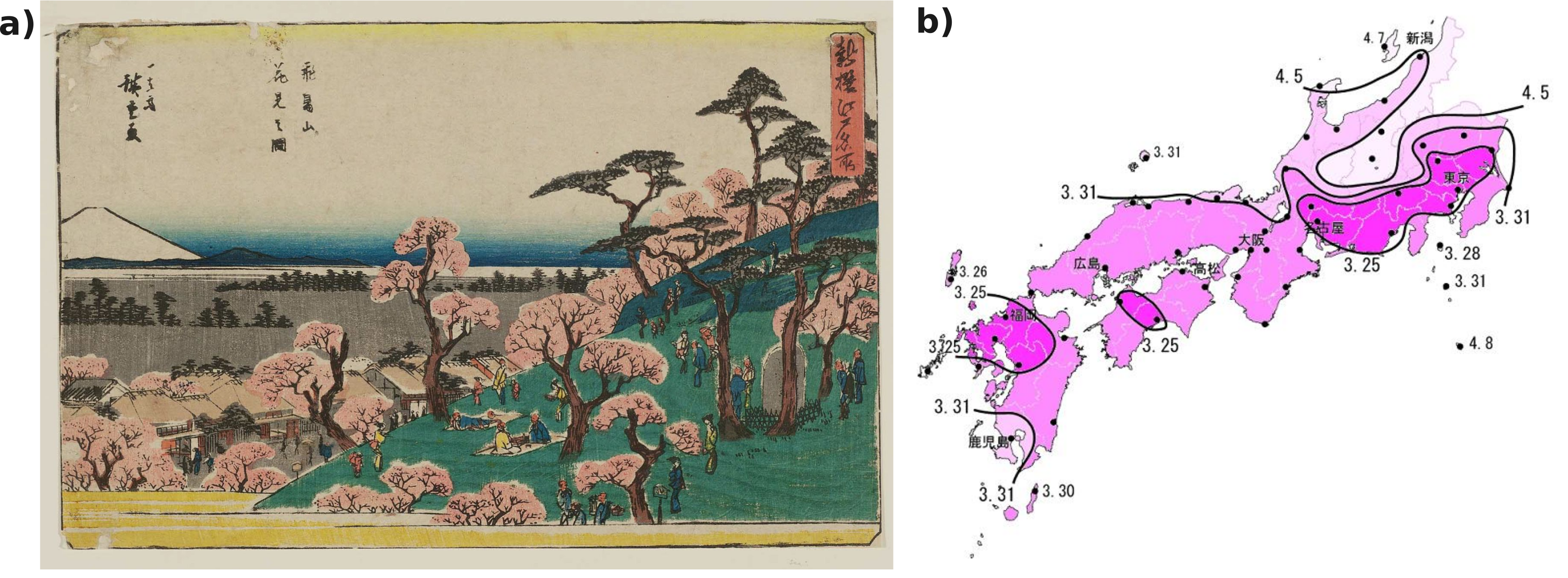}
\caption{In \textbf{a)}, picture called \textit{Asukayama hanami no zu} (Cherry-blossom viewing at Asuka hill) by Utagawa Hiroshige, dated about 1831. In \textbf{b)}, forecasting of the dates of blossoming across Japan in 2007. The numbers represent the month and day of the blossoming, being the darker areas the ones where blossoming is expected earlier. Source: Wikimedia Commons.} \label{fig_hanami}
\end{center}
\end{figure}

The bloom of flowers in spring, the arrival of the first migratory birds, and the shedding of leaves in fall are all simple yet beautiful examples of phenological events that undeniably shape, since ancient ages, our perception and narration of time, especially the passage of the seasons. Our language and culture are full of references to this sort of phenomenon and their timing, from words like `late bloomer' to Aristotle's famous phrase `\textit{one swallow does not make a summer}'. Naturally, hence, we can find as well ancient instances of documented phenological patterns, not only related to agricultural needs but also in cases in which their timing marked the date of religious or traditional festivities. A fascinating illustration is the viewing of cherry blossoms in Japan, known as \textit{hanami}, a celebration that started in the eighth century as an elitist ceremony and gradually became a general festivity. Nowadays, it is so popular that the cherry blossom is nightly forecasted in Japan, and the advance of the front of blossom across the country -the so-called \textit{sakura zensen}- is keenly followed (see Fig.~\ref{fig_hanami}). As a result, it is possible to find extensive datasets on the phenology of various species of the cherry tree over the years --mainly the \textit{Prunus serrulata}-, comprising more than seven centuries of information on blossoming dates\cite{aono2008phenological}. In fact, this kind of sequential data has been used to reconstruct springtime temperatures of undocumented old periods. Similarly, analogous studies have been carried out using alternative species' phenology around the globe, such as the grape plant to produce wine in France\cite{chuine2004grape}. 

While these datasets were constructed by relying, basically, on visual and small-scale observations, during the last decades technological advances have permitted a multiplication and diversification of the means to measure the phenology of both plants and animals. Such methodologies range from citizen science projects where participants collect and share individual observations of phenology in their local area\cite{havens2013citizen}, to the use of sophisticated remote sensing tools based on satellite data\cite{zhang2003monitoring}. A particularly ingenious case among these novel approaches is the work by\cite{graham2010public}, in which they process the images recorded by public cameras connected to the Internet -primarily related to traffic surveillance or national parks- to gather information on vegetation phenology across North America, exploiting a free source of glances of trees and plants unintendedly caught by the cameras. Nurtured by these methodological advances as well as spurred by the investigations on how climate change influences phenology\cite{memmott2007global,hegland2009does}, nowadays we face an undeniable increase in the quantity and the quality of documented phenology. 

Paradoxically, though, this copious amount of phenological data comes at relative help when trying to better understand mutualism over time. This is due to the fact that public empirical information on both the network of interactions \textit{and} the timing of their mutualistic activity is rather scarce.  Consequently, previous studies examining the effect of phenology on mutualistic communities either focus on a very reduced number of highly-resolved networks\cite{olesen2008temporal,caradonna2017interaction,ramos2018phenology}, either they succeed in keeping the big numbers of phenological data at the expense of roughly approximating the patterns of mutualistic relationships --that is, by neglecting the real complex networks of interactions\cite{duchenne2020phenological}. A third approach, yet, to address the lack of data, is to explore instead synthetic models either for phenology or for the network of interactions\cite{kallimanis2009plant,encinas2012phenology,duchenne2021phenological}. \LH{In this work we chose to use the latter in order to overcome the problem of data scarcity.}\removedLH{In what follows, these limitations will condition us as well.}

\subsection{Ecological and evolutionary determinants of phenology}

Although phenology certainly plays a part in different types of ecological mutualism, in this work we will focus on its implications for plant-pollinator communities, both because it is a paradigmatic example in which seasonality is strongly marked and also because, despite the aforementioned data limitations, the phenology of plants and pollinators is, in general, better documented than that of other mutualistic species\cite{rafferty2015phenological}. 

The study of the phenology involves all temporal aspects of species' life cycle, regarding on the one hand the timing of their various stages of development, i.e. egg/seed, larvae, adult, etc.; and on the other hand the onset and duration of different biological processes such as flowering, germination, pollination, leaf shedding, etc. Such timing is determined by a myriad of factors, similar to what occurs with mutualistic interactions in general\cite{vazquez2009evaluating}. Indeed, both biotic and abiotic forces shape the phenology of species, which, in addition, is thought to be subject to evolutionary change\cite{rathcke1985phenological, forrest2010toward}. To complicate things further, several sources of intra-species phenological variability occur simultaneously: inter-annual\cite{olesen2008temporal,cirtwill2018between} --that is, among different seasons--, geographical\cite{post2018acceleration} --for the same species but on diverse sites-- and last but not least, individual, i.e. among different individuals of the same population, even those coexisting on the same site and at the same season\cite{forrest2010toward}. 

\begin{figure}[h!]
\begin{center}
  \includegraphics[width=0.6\linewidth]{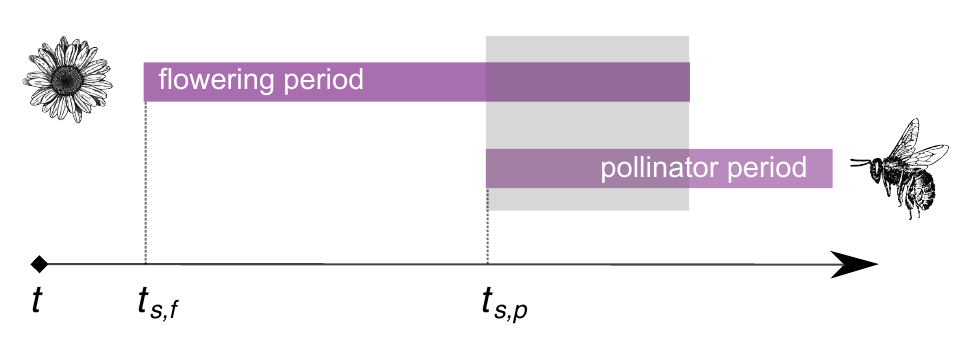}
\caption{Schema of the phenology of a plant and a pollinator, represented by their \textit{active} states. Each species is defined by its period and its starting date, called here $t_{s,f}$ for the flowering plant and  $t_{s,p}$ for the pollinator. The portion of time during which the two species overlap and can actually interact is highlighted by the gray square.} \label{fig_activity}
\end{center}
\end{figure}

All in all, from a statistical viewpoint a few general patterns have been identified. In particular, we will focus on two fundamental phenological quantities of plant-pollinator systems that are well characterized: the \textit{starting dates}, that is, the time at which the flowering --in the case of plants-- or the pollination --for animals-- begins, and secondly the so-called \textit{periods}, that is, the length or duration of this active state, during which mutualistic interactions are possible. Certainly, this selection of phenological indicators is far from being exhaustive, and other works have modeled the different stages of plants and pollinators in more realistic detail\cite{ramos2018phenology}. In our approach we will approximate this complex landscape of temporal variability by focusing only on the adult phase of the species, and moreover, reducing the whole possible set of biological processes and stages to a couple of states: \textit{active}, namely when the species could potentially hold a mutualistic relationship --i.e. pollinate or be pollinated--, or \textit{inactive}. This sort of simplification is not novel and other works have adopted analogous approaches\cite{memmott2007global,encinas2012phenology,burkle2013plant}. 
In Fig.~\ref{fig_activity} we depict this schematic representation of a hypothetical plant and its pollinator.
 
Regarding the periods of activity, in general, both plants' and pollinators' periods tend to follow right-skewed distributions, with a small number of species exhibiting a long phenophase while a large number of species are active only during a short time\cite{bawa2003relationships,kallimanis2009plant}. Particularly, \LH{Kallimanis et al.}\cite{kallimanis2009plant} examined the statistical distribution of period lengths in a Mediterranean scrub community observed over four years by \LH{Petanidou et al.}\cite{petanidou1995constraints} and concluded that the distribution of pollinator's periods could be fitted by a decreasing exponential, while plants' followed a lognormal distribution. On the other hand, \removedLH{identified} a lognormal shape for both plants and pollinators' periods and a normal distribution for plants in one of the years of their observation \LH{has been identified in a plant-pollinator system in the Arctic}\cite{olesen2008temporal}.

In what concerns the onset of flowering and pollinating activity, a complementary measure that is often used is the \textit{middle date}. In this sense, evidence suggests that middle dates tend to be relatively synchronized\cite{rathcke1985phenological,tebar2004flowering}. For example, both \LH{in Ref.}\cite{olesen2008temporal} and\cite{bawa2003relationships} \LH{it was} observed that most active periods temporally coexist at a peak, probably due to a similar reaction to a common set of physicochemical stimuli such as temperature, photo-period, humidity, etc\cite{rathcke1985phenological}. In statistical terms, this would correspond to a scenario where the middle dates are relatively clustered, as modeled for instance\removedLH{by} \LH{in Ref.}\cite{kallimanis2009plant} using a normal distribution. At the same time, genetic factors seem to play an important role in determining the timing of flowering or pollinating activity, which contributes to explaining the heterogeneity of starting dates. On the other hand, it has also been hypothesized that species may spread along the season in order to minimize competition. Although the evidence supporting this hypothesis is controversial\cite{rathcke1985phenological}, some authors have proposed that such minimization could still occur within a temporal range determined by genetic and environmental constraints\cite{kochmer1986constraints}. 

This short introduction illustrates the complexity of the temporal dimension of plant-pollinator communities. Indeed, temporal variability not only appears at different scales, from days to decades and from individuals to species, but it is furthermore regulated by a multiplicity of factors. Moreover, the relevance of phenology goes beyond the individual characterization of species' traits, since it regulates as well the ecological interactions within the community, affecting both their occurrence and intensity. Understanding how this may modify the structure of ecological interactions beyond the limits of the aggregated paradigm is crucial if we aim to produce a refined depiction of natural systems. In order to do so, in the next section we define a framework to model the effects of phenology on mutualistic networks using real data, whose outcomes we will explore in the results section.

\section{Definitions and Models}

Along this section we introduce the tools and models to explore the effects of phenology on a plant-pollinator community: first, we define the mutualistic and competitive networks of interactions; second, we describe how we may account for empirical phenology using overlaps and a coarse-grained temporal representation of networks; finally, we present a set of basic toy-models to produce synthetic configurations of phenology using both mechanistic and statistical principles.

\subsection{Mutualistic and competitive network}

In order to assess the effects of phenology on the plant-pollinator community, we will not only consider the mutualistic relationship among different species but also indirect interactions like competition for shared mutualistic resources\cite{jones2012fundamental}, which naturally emerge among species of the same kind. The negative effects of these antagonistic interactions are known to coexist with mutualism, leading to a trade-off between costs and benefits\cite{gracia2018joint,bronstein2001costs}. 

Therefore, we start by defining the biadjacency matrix of mutualistic interactions $B_{i,k}$ of size $N_P \times N_A$, where $N_P$ is the number of plants and $N_A$ is the number of animals, such that:

\begin{eqnarray}
\small
\mbox{ if } B_{i,k} = 1, \mbox{ then plant } i \mbox{ is pollinated by animal } k;\nonumber\\
\mbox{ if } B_{i,k} = 0, \mbox{ then species } i \mbox{ and } k \mbox{ do not share any mutualistic interaction.}\nonumber\\ 
\label{eq_mutualistic_interactions}
\end{eqnarray}

In order to be able to calculate the consequence of temporal variability also among competitors, we need to infer the network of competitive interactions based on shared mutualistic resources. We do so following\removedLH{the proposal by} \LH{Ref.}\cite{gracia2018joint}, namely, projecting the empirical biadjacency matrix of mutualistic interactions $B_{i,k}$ into the subspace of intra-guild interactions such that:

\begin{eqnarray}
\small
\mbox{ if } B_{i,k} B_{j,k} = 1, \mbox{ then plants } i \mbox{ and } j \mbox{ compete for the pollinating services of animal } k ;\nonumber \\
\noindent \mbox{ if } B_{i,k} B_{j,k} = 0,   \mbox{ then } i \mbox{ and } j \mbox{ do not compete for the resources of pollinator } k,\nonumber\\
\label{eq_competitive_interactions}
\end{eqnarray}
which leads to a competitive network between all pairs of plant species $i$ and $j$. Analogously, we can define the competitive interactions among two pollinator species $k$ and $l$ as follows:

\begin{eqnarray}
\small
\mbox{ if }  B_{i,k} B_{i,l} = 1,  \mbox{ then pollinators } k \mbox{ and } l \mbox{ compete for the flower resources of plant } i ; \nonumber\\
\mbox{ if } B_{i,k} B_{i,l} = 0, \mbox{ then } k \mbox{ and } l \mbox{ do not compete for the resources of plant } i.\nonumber\\
\end{eqnarray}

Altogether, this provides a bilayered depiction of the ecological community, such that the observed network of pollinating contacts not only mediates the explicit mutualistic interactions, but also the implicit, competitive relationships among species of the same guild\cite{gracia2018joint}.   

\subsection{Empirical overlaps}

\begin{figure}[h!]
\begin{center}
  \includegraphics[width=\linewidth]{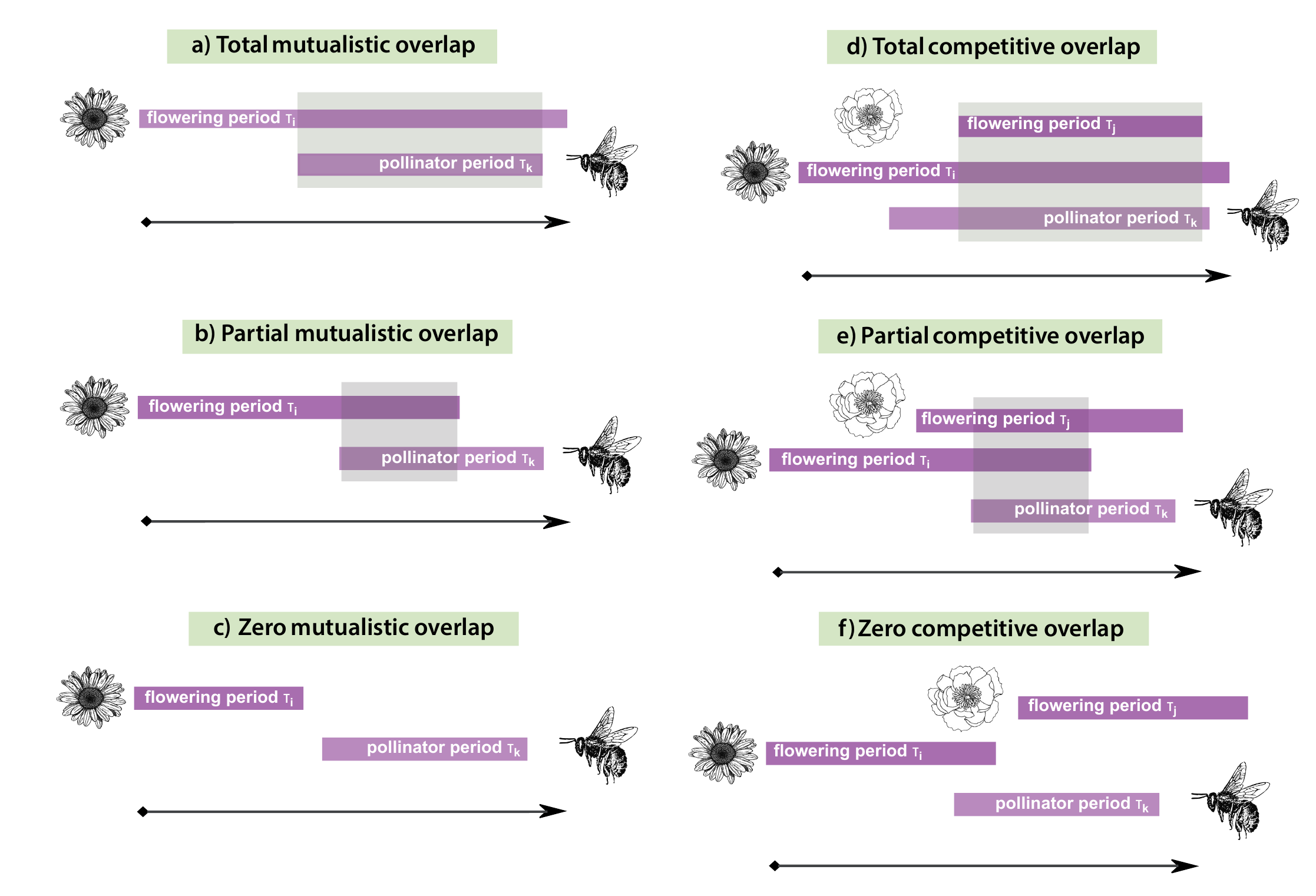}
\caption{Diversity of phenological configurations giving rise to different types of mutualistic and competitive overlap. The panels on the left represent the mutualistic overlap between the pollinator species $k$ and the plant species $i$, from the viewpoint of the pollinator species $k$. In \textbf{a)} there is full mutualistic overlap, in \textbf{b)} there is only partial overlap, and in \textbf{c)} there is no overlap at all. Right Panels depict the competitive overlap between three species, particularly two plants, $j$ and $i$, competing for a shared pollinator species $k$. In this case, the overlap is referred to the first species, $j$. In \textbf{d)} the competitive overlap is full, in \textbf{e)} there is only partial overlap, and in \textbf{f)} there is no overlap at all between the competitors.} \label{fig_overlaps}
\end{center}
\end{figure}

The most straightforward consequence of introducing phenology into an aggregated network is the modification of the amount of temporal coexistence --the so-called \textit{overlap}-- among species. Indeed, we find a continuum of possible scenarios ranging from full concurrence to absence of overlap, as shown in Fig.~\ref{fig_overlaps}. These overlaps concern either two species in the mutualistic case or three species --two species of the same kind and the shared resource-- when considering the competition. 

Within this framework, we can introduce a set of phenological coefficients $\{\Phi\}$ and $\{\Omega\}$ to quantify the effect of the phenological overlap on, respectively, the mutualistic and the competitive interactions. In the particular case of a plant species $i$ interacting mutualistically with a pollinator $k$ and competitively with another plant $j$, we define these coefficients as follows:

\begin{equation} \label{eq_phi}
\Phi^P_{ik} = \frac{\tau_{ik}}{\tau_i},
\end{equation} 

\begin{equation} \label{eq_omega}
\Omega^P_{ijk}= \frac{\tau_{ijk}}{\tau_{i}},
\end{equation}

\noindent where $\tau_{i}$ stands for the period of flowering of the plant species $i$, $\tau_{ik}$ represents the temporal overlap between the \LH{period of the} plant species $i$ and \LH{that of} its pollinator $k$ (see Fig.~\ref{fig_overlaps} a-c for a graphical representation), and finally $\tau_{ijk}$ stands for the overlap between plants species $i$ and $j$ and their pollinator $k$ (see Fig.~\ref{fig_overlaps} d-g for an example). An analogous set of coefficients can be drawn for the pollinators. 

Note that, as can be seen in Eqs.~\ref{eq_phi}-\ref{eq_omega}, the overlaps are pondered by the period of activity of the species --in the example above, the plant $i$--, which means that each coefficient is always referred to a certain plant or animal species. This particular normalization implies, moreover, that the effect of the temporal overlap is non-symmetric among the interacting partners and hence, in general, $\Phi^P_{ik} \neq \Phi^A_{ki}$.

\subsection{Coarse-grained temporal sequence of networks}

In order to better understand how the perceived structure of the network changes if we take the empirical phenology into account, we construct as well a discrete temporal sequence of networks. Here, each element corresponds to a different \textit{snapshot} of the system, taken at a different moment of the season. In particular, given that the degree of detail of the empirical phenologies under study is narrowed to \textit{days}, we construct a set of daily networks. That is, each network represents the mutualistic interactions observed among plants and pollinators on a given day. 

To construct this sequence of networks, we use the empirical information on the starting and ending dates of activity of plants and pollinators, then remove from the daily network certain mutualistic links among species whenever they do not coincide in time -- i.e., when their phenological overlap, as defined in Eq.~\ref{eq_phi}, is zero. Additionally, we remove inactive species as well as active species with no interactions --i.e. zero degree.

\begin{figure}[h!]
\begin{center}
  \includegraphics[width=\columnwidth]{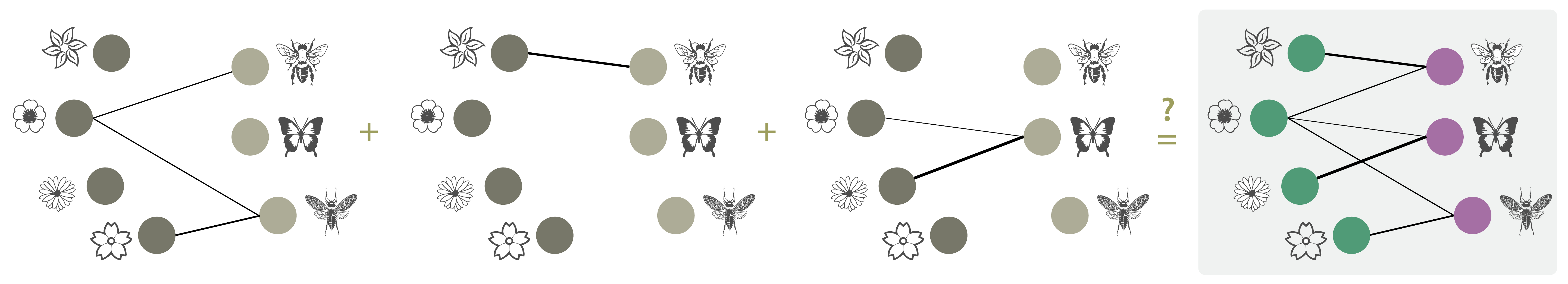}
\caption{Schematic representation of the coarse-grained description of temporal variation. Each network in the grey scale corresponds to a different day of the season, where certain species and interactions might be absent. Summing up these interactions produces the aggregated network, depicted in color, where the information on the turnover of species and interactions is lost. Despite this aggregation process is indeed quite simple, a pertinent question is whether the structure and dynamics of the temporal snapshots are comparable to those of the aggregated network.} \label{fig_temporal_networks}
\end{center}
\end{figure}

On the whole, this provides a coarse-grained description of the community as shown in Fig.~\ref{fig_temporal_networks}, that, even if it is not a pure temporal network, it does account to a great extent for the intra-annual temporal variation, moving beyond the static network formalism.

\subsection{Synthetic toy-models}

In order to evaluate how our results depend on the details of the empirical dataset, we develop a group of toy models that, under different assumptions, produce synthetic configurations of phenology compatible with a given mutualistic network. As aforementioned, in our framework, the phenology of each species is defined by two elements: the starting date and the period of activity. Here, we will focus on determining the starting dates\removedLH{and thus preserve} \LH{while preserving} the empirical periods. Therefore, the starting dates will be treated as parameters to be set by the synthetic model, under the condition that mutualistic partners share a non-zero mutualistic overlap. This requirement ensures that the binary mutualistic network is kept equal to the observed one, constituting a formalism that focuses explicitly on modeling the phenology.

Before entering into further details about the different synthetic models, let us briefly discuss the particularities and consequences of preserving the network of interactions. As aforementioned, we will construct a model under the requirement that the phenological overlap among mutualistic partners is comprised in the range $(0,1]$, excluding the zero in order to ensure that the two species still interact. This means that, strictly speaking, we will only keep the \textit{binary} mutualistic network. In other words, the general idea consists in constructing a synthetic phenology that is \textit{compatible} with the given network treated in binary terms.

In this line of thought, the constraint that mutualistic partners share non-zero phenological overlap may be written as an inequality for each species' starting date. For instance, let us take a plant \LH{species} $i$ with period $p_{i}^P$ and unknown starting date $t_{0,i}^P$, \LH{which interacts}\removedLH{who is interacting} with an animal \LH{species} $j$ with period $p_{j}^A$ and an unknown starting date $t_{0,j}^A$. In order to ensure that their periods overlap, it must be fulfilled that:

\begin{equation}
t_{0,j}^A - p_{i}^P + 1 \, \leq \, t_{0,i}^P \, \leq \, t_{0,j}^A + p_j^A -1 \, , \label{ineq_t0}
\end{equation} 
where we have assumed a daily coarse-graining, such that the minimal overlapping unit is one day. In order to ensure that all interacting species share phenological overlap, we may write an inequality analog to Eq.~\ref{ineq_t0} per \LH{each} link in the bipartite network. Then, if we take the starting dates as unknown variables, we can view the problem of constraining the starting dates as a multivariate system of linear inequalities composed by $N$ unknown variables and $L$ inequalities (where $N$ and $L$ are respectively the number of nodes and the number of links of the network). Note that both $\{t_{0}^P\}$ and $\{t_{0}^A\}$ are unknown, which means that the system of inequalities is coupled.

Furthermore, in order to ensure that the resulting distribution is realistic, we may introduce an additional set of constraints. In particular, we impose a lower and upper bound for the starting dates of activity, which ensures that the community exhibits some seasonality, a characteristic that has been empirically observed\cite{rathcke1985phenological,kallimanis2009plant}. If we represent by $l$ the lower bound and by $u$ the upper bound delimiting the season, then the condition on the starting date of an arbitrary species $i$ can be written as:

\begin{equation}
l \, \leq  \, t_{0,i} \, \leq \, u - p_i +1 . \label{ineeq_t0_bounds}
\end{equation} 

Overall, Eqs.~\ref{ineq_t0}-\ref{ineeq_t0_bounds} represent a coupled system of linear inequalities, where the parameters to be determined are the starting dates. Although a problem of this kind might not generally have a feasible solution, in our particular case the existence of some trivial solutions --e.g. the same starting dates for all species-- warrants that the system is consistent. 

Within this framework, we present a group of approaches that, under different assumptions, permit determining the set of synthetic starting dates while fulfilling the conditions in Eqs.~\ref{ineq_t0}-\ref{ineeq_t0_bounds}. In particular, we explore four possible scenarios, corresponding to different fundamental hypotheses about the main factor governing the species' phenology, namely: \textit{(i)} species' periods tend to be synchronized, \textit{(ii)} species seek to organize their activity as to minimize their intra-guild competition for mutualistic resources, \textit{(iii)} species periods get maximally dispersed along the season, \textit{(iv)} in the absence of further information, the starting dates can be determined using a minimal model of maximum entropy. 

As can be observed, each of these assumptions relies on a single dominant factor. Indeed, hypothesis \textit{(i)} and \textit{(ii)} are mechanistic, in the sense that they place the emphasis on environmental and ecological drivers of the phenology. Instead, the scenarios \textit{(iii)} and \textit{(iv)} function as non-mechanistic models based on statistical properties of the distribution of starting times, regarding either their dispersion or their entropy. Of course, since biotic and abiotic forces do not act separately, one could construct a more complex model where various of these factors operate simultaneously. In this sense, these synthetic models constitute a first-order approach, which\removedLH{nonetheless, as we will see, yields some interesting results} \LH{helps to understand the consequences of each of these hypotheses concerning the phenology}. In what follows, we define each of these models one by one and describe how we numerically implemented them.

\subsubsection{Synchronized phenology}

This model is based on the assumption that the level of synchronization in the community is high, hence most species' activity temporally coincides with the peak of the season. As we reviewed in the introduction, empirical evidence suggests that seasonal systems usually exhibit non-negligible synchronization, in the sense that the flowering and pollination of species tend to occur simultaneously. Several hypotheses have been proposed to account for this pattern, from the presence of abiotic pressures such as the temperature, the precipitation, or the photoperiod --disregarding genetic differences or biotic constraints--\cite{forrest2010toward}, to the existence of facilitative interaction among concurrent species, which obtain an ecological benefit of overlapping their periods of activity. All in all, it is still not clear how the mentioned factors interplay among themselves and with other possible forces to give rise to the observed heterogeneity in starting dates\cite{forrest2010toward}.

In order to attempt to reproduce this pattern, we construct a model where species' periods are approximately centered. To do so, we set the middle dates around the peak of the season and then slightly perturb them, following the numerical procedure detailed in the Supplementary Material. This leads to phenological configurations where species are considerably --but not perfectly-- synchronized. Indeed, introducing this imperfection is crucial for two reasons: first, to account for the aforementioned heterogeneity in starting and middle dates of activity; second, because a perfectly synchronized system is trivial, in the sense that all species fully overlap, recovering hence the aggregated case.

\subsubsection{Minimization of the intra-guild competition}

This second model is built upon the hypothesis that the periods of the species are located along the season so as to avoid, as far as possible, overlapping with other species that exploit a common mutualistic resource --i.e. pollinating services in the case of plants or flower availability for pollinators. As discussed above, the evidence for this type of mechanism is controversial, and, when it does occur, its effect is most probably subordinated to other genetic and environmental factors. Nevertheless, we will still explore the consequences of this assumption, given that it is the natural counterexample of the synchronized scenario.

In order to construct a phenological configuration that minimizes the competition among species, we implement a global search algorithm that optimizes an objective function under the constraints in Eqs.~\ref{ineq_t0}-\ref{ineeq_t0_bounds} (see the Supplementary Material for the numerical details). In this particular model, the objective function corresponds to the total competitive phenological overlap among species.

\subsubsection{Maximization of the variance}

We now turn our attention to models that focus directly on the statistical properties of the system's phenology. As a first approach, we propose a model based on maximizing the dispersal of the periods along the season. In particular, we maximize the variance of the middle dates of activity. Indeed, if we define the middle time $t_M$ of a species $i$ as:

\begin{equation}
t_{M,i} \, = \, t_{0,i} \, + \, \frac{p_{i} - 1}{2} \, , \label{eq_tm}
\end{equation}
then its variance Var$(t_{M})$ is an approximate measure of the dispersion of the activity of both plants and pollinators along the season. Once again, the maximization of this quantity is done under the condition that mutualistic interactions present in the aggregated network still occur. The numerical procedure to do so is explained in the Supplementary Material.

\subsubsection{Maximization of the entropy}

Following a similar line of thought, we propose a second model to produce phenological configurations that is based on their statistical properties. In particular, we refine the former approach by characterizing the distribution with a more sophisticated and robust indicator, namely, entropy. This is a plausible alternative to maximizing the variance which may especially outperform the latter when the distribution is multimodal. 

Let us define the entropy within the constructed framework for describing phenology. We start by defining a random variable $x_i$ as the number of species whose middle date of activity $t_m$ takes place at a given date $t_i$. Then, its corresponding probability $p(x_i)$ will be:

\begin{equation}
p(x_i) \,  = \, \frac{\mbox{number of species with } t_m = t_i}{\mbox{total number of species}} .
\end{equation} 

Given this random variable, it is possible to define its Shannon-Gibbs entropy as follows:

\begin{equation}
S\, = \, - \sum_i \, p(x_i) \, \mbox{ln } p(x_i) . \label{phen_entropy}
\end{equation} 

It is straightforward to see that in a scenario of perfect synchronization, the middle dates of activity are the same for all species. That is, if we name such a date by the index $k$, then $p(x_k)=1$, and the information entropy is zero. 

Maximizing the entropy defined in Eq.~\ref{phen_entropy} under the constraints imposed by the network of interactions results in a maximum entropy model, that randomizes the distribution of periods of the species yet using only the minimal amount of necessary information. The particular details of how this model was implemented can be found in the Supplementary Material.

\section{Datasets}

In order to assess the consequences of phenology on the perceived network structure, we analyzed two public datasets\removedLH{by} \cite{burkle2013plant, kantsa2018disentangling}\removedLH{and} which contain observations of mutualistic interactions in plant-pollinator systems together with their phenology. We detail the peculiarities of each dataset below.

\subsection{The Illinois dataset}

\removedLH{The} \LH{This dataset}\removedLH{gathered by} \cite{burkle2013plant} corresponds to a set of woodland sites in Carlinville, Illinois (USA), observed during the springs of 2009 and 2010 from March to May. The observed network of mutualistic interactions contained originally 26 spring-blooming herbaceous plants and 54 bee species. However, 2 plant species did not exhibit any interaction in the empirical network, so we obtained eventually a $24 \times 54$ network.  

The dataset also included the observed phenology, specifically the starting and ending dates of the active stages of both plants and pollinators. Using those, we extracted as well their periods of activity and the middle dates.

\subsection{The Lesvos Island dataset}

This dataset\removedLH{recorded by} is based on a two-year study conducted in Aglios Stefanos, Lesvos Island (Greece) during the springs of 2011 and 2012 from April to July\cite{kantsa2018disentangling}.\removedLH{The published dataset} \LH{It} includes information on both the network of mutualistic interactions and the phenology. Aggregated over the two years, the resulting network is composed of 41 plant species and 168 pollinators. 

Remarkably, a large portion of the pollinator species is present in only one of the seasons. In particular, out of the total 168 pollinator species observed over the two years, only 67 species ($39.9\%$) were persistently found in both seasons, while 53 species ($31.5\%$) were observed just during the first year, and 48 species ($28.6\%$) just during the second year of observation. This implies that not only there is an important turnover of interactions among the two consecutive years, but also there is a considerable species turnover, as suggested as well by previous studies\cite{olesen2008temporal,chacoff2018interaction}.

In order to obtain the network of interactions specific to each year, we removed those interactions present in the aggregated network\removedLH{that, after including the corresponding information on phenology, occur among species} whose periods of activity do not overlap \LH{according to the information concerning phenology}. In such cases, it is clear that the interaction was not possible due to the lack of phenological overlap and hence we can remove it from the yearly network. After doing so for each season, we removed as well those species that no longer hold mutualistic interactions --i.e. they have degree zero. As a result, both year-specific network sizes' are reduced to 34 plant species and 113 pollinators in the first year and 38 plant species and 104 pollinators in the second year of observations.

\section{Results}

\subsection{Characterization of the empirical datasets}

We start by examining the distribution of the duration of the activity of plants and pollinators. In particular, we fit a variety of functional forms on the cumulative distributions using a maximum likelihood estimation. Then, we test the quality of the fit by performing a Kolmogorov-Smirnov test by bootstrapping, as explained in detail in the Supplementary Material. 

In detail, we test three different functional forms: lognormal, exponential, and beta function. As aforementioned, the first two distribution types have been claimed to correctly describe some empirical observations, while we introduce the latter as a generic fitting. The results of the p-value for the KS-test are reported in Table~\ref{table_fit_periods}, where we highlight in bold the good-quality fits. As can be seen, the Illinois dataset is better fitted by the beta function, while in the Lesvos Island dataset, we can fit a lognormal to the distribution of plants' periods but we found no unique fit for the pollinators. In general terms, the distribution of periods of the Illinois dataset is less heterogeneous, while the Lesvos Island distributions are clearly righter-skewed. This is especially notorious for the periods of the pollinators, among which we find a large proportion of species with very short periods. 

\begin{center} 
\begin{table}[h]
\caption{Results of fitting different functional forms, disentangled into plants and pollinators. We show the p-value of the fit, and highlight in bold the good quality fits, that is, those which do not significantly differ from the estimated distribution (p-value $> 0.05$). The p-values are obtained by performing a Kolmogorov-Smirnov test between the fitted distribution and the empirical sample, then comparing it to the corresponding K-S distribution sampled by bootstrap as explained in the Supplementary Material.}
\center
\begin{tabular}{c c c c c c c} 
\hline
\hline
Dataset &\multicolumn{2}{c}{ Lognormal fit} & \multicolumn{2}{c}{Exponential fit} & \multicolumn{2}{c}{Beta function fit}\\ 
\hline
 & Plants & Pollinators & Plants & Pollinators & Plants & Pollinators \\
\hline 
Illinois &  0.021  & 0.018 &  0.001   &  0.001  &  \textbf{0.63}  &  \textbf{0.18}  \\ 
Lesvos Island 1st year &  \textbf{0.67}  &   0.014  & 0.029  &  0.001  & \textbf{0.41}  & 0.001 \\  
Lesvos Island 2nd year & \textbf{0.50}  &  0.001   &   0.018  & 0.001 & 0.02 & 0.001 \\
\hline  
\hline
\end{tabular}
\label{table_fit_periods}
\end{table}
\end{center} 

\subsection{Temporal variation of the structure}

\begin{figure}[h!]
\begin{center}
  \includegraphics[width=\linewidth]{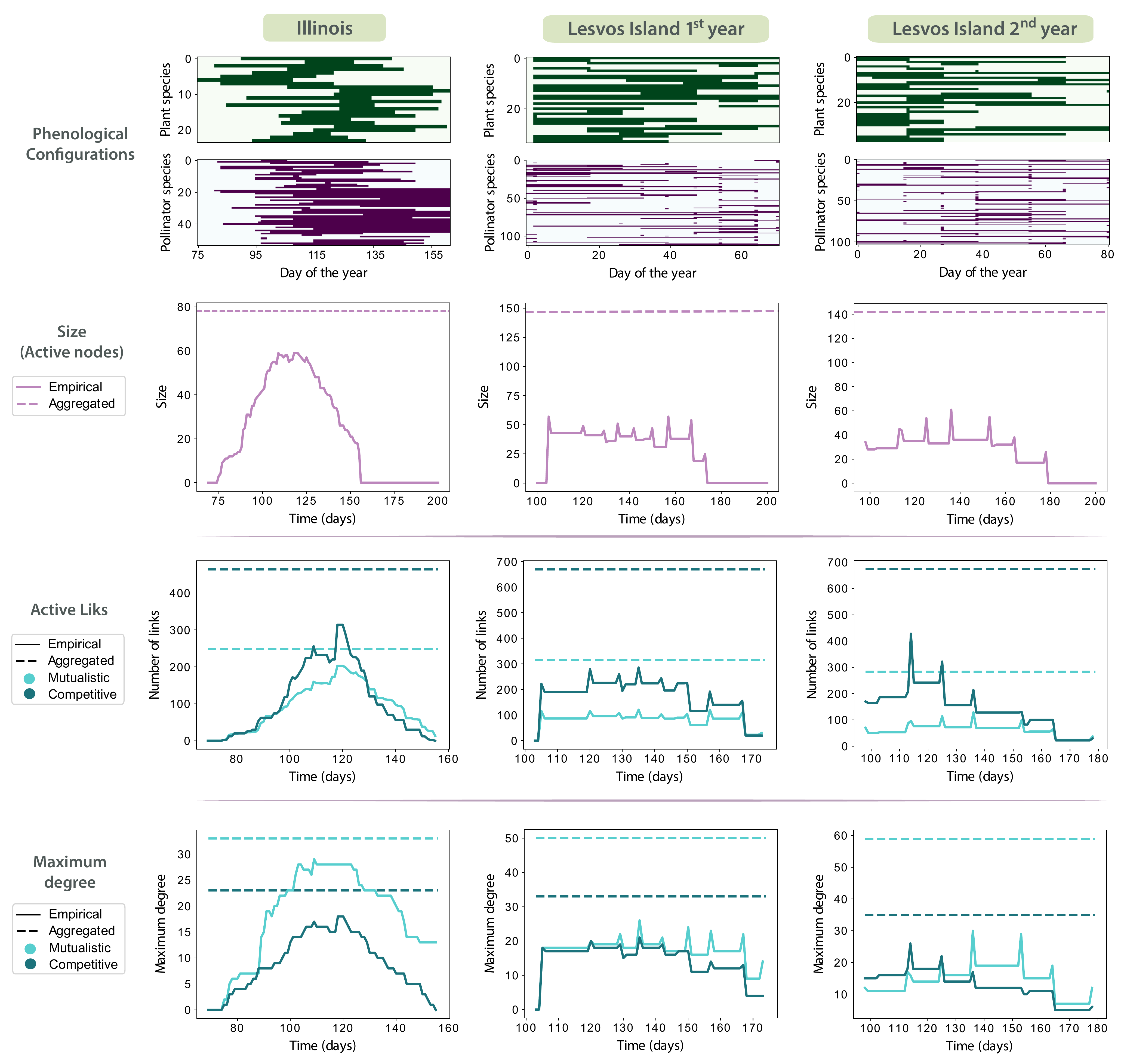}
\caption{Phenological configurations and temporal evolution of three fundamental structural features. On the top, a schema of the temporal arrangement of the periods of activity of plant species (green) and pollinators (purple) along the season. Below, is the temporal variation of three basic structural parameters, measured on a daily basis: number of active nodes, active links, and maximum degree. Each column corresponds to a different dataset as indicated on the top of the figure.} \label{fig_temporal_structure}
\end{center}
\end{figure}

Now that we have characterized the empirical phenology of the two datasets, we turn our attention to the question: how does the perceived structure of the ecological network change if we take phenological information into account? In particular, we measure a set of fundamental structural features for the different coarse-grained networks of the daily sequence: the number of active nodes, the number of active links, and the maximum degree. We do so both for the mutualistic and the competitive network. Fig.~\ref{fig_temporal_structure} provides a summary of the variation of these properties over time, for each of the empirical networks in our dataset. 

To begin with, our results show that in both cases the temporal measures are significantly smaller than their aggregated counterparts. However, we observe a non-negligible quantitative difference among datasets: in the Illinois dataset, the maximum of the peak represents around the 75-80$\%$ of the aggregated counterpart, whereas in both instances of the Lesvos Island dataset, this quantity descends to less than 30-40$\%$. Additionally, the comparison of the results for the two systems shows that the variation of their network structural properties exhibits a strikingly different pattern over the season. Indeed, while in the Illinois dataset, the three quantities show a marked peak approximately at the middle of the season, in the Lesvos Island dataset they stay relatively stable, showing an almost flat shape.

\subsection{Comparison with synthetic models} 

We now proceed to test the performance of the different generative models of synthetic phenologies proposed above. To do so, we will adopt as input the observed network of interactions --which is to be preserved--, and the empirical distribution of periods. The latter will allow us to focus on the quality of the methodology to produce the starting dates alone, without introducing additional sources of noise.

In order to examine the adequacy of each model to reproduce the empirical phenological patterns, we start by extracting the synthetic starting dates corresponding to each of our three datasets (see the Supplementary Material for the explanation of the numerical methods). Then, we characterize the synthetic phenologies with the methods and definitions described in section 2. That is, we calculate, on the one hand, the distribution of phenological coefficients defined in Eqs.~\ref{eq_phi}-\ref{eq_omega}, and, on the other hand, we couple the phenology to the network structure and compute the change of some basic structural features along time. Along the rest of this section, we will explain and discuss the results of comparing these synthetic measures with the empirical observations.

\subsubsection{Distribution of overlaps}

We first study the distribution of the phenological coefficients defined in Eqs.~\ref{eq_phi}-\ref{eq_omega}. In order to compare the empirical and the synthetic distributions, we perform two different statistical measures: \textit{(a)} a two-sample Kolmogorov-Smirnov test and \textit{(b)} the calculation of the Kullback-Leibler divergence (see the Supplementary Material for the numerical details).

Using \textit{(a)}, we analyzed separately the competitive and mutualistic overlaps on the one hand, and the distributions for plants and for pollinators on the other hand, as summarized in Table~\ref{table_ks_twosample}. In particular, for the Illinois dataset, the null hypothesis that both samples come from the same distribution is rejected for the models which minimize the competition and maximize the variance, which is hence incompatible with the empirical data. The model with centered times and maximum entropy is generally compatible with the observed phenology, except for the distribution of coefficients associated with the competition among pollinators. Instead, for the Lesvos Island dataset, the four models are compatible with the data.

\begin{center} 
\begin{table}[h]
\caption{Results of the K-S two sample test among the synthetic distribution and the empirical distribution. We show the KS distance between the two samples, and highlight in bold the cases in which the null hypothesis is rejected, that is, those where the two samples do significantly differ (p-value $< 0.05$). The smaller the KS distance, the smaller the discrepancy between the two samples. The numerical implementation to calculate these distances is described on the Supplementary Material.}
\centering
\begin{tabular}{c c c c c} 
\hline
\hline
\textbf{Model} /Dataset &\multicolumn{2}{c}{Mutualistic overlap} & \multicolumn{2}{c}{Competitive overlap}\\ 
\hline
\hline
\textbf{Synchronized} & Plants & Pollinators & Plants & Pollinators  \\
\hline 
Illinois & $\;$ 0.052 $\quad$ & $\quad$ 0.051 $\quad$ & $\quad$ 0.008  $\quad$ & $\quad$ \textbf{0.021} $\quad$ \\ 
Lesvos Island 1st year & $\;$ 0.015 $\quad$ &  $\quad$ 0.010 $\quad$ & $\quad$ 0.002 $\quad$ & $\quad$ 0.002 $\quad$\\  
Lesvos Island 2nd year & $\;$ 0.009 $\quad$ &  0.009  & $\quad$ 0.001 $\quad$ & 0.0017 \\
\hline  
\hline
\textbf{Minimum competition} & Plants & Pollinators & Plants & Pollinators  \\
\hline 
Illinois & $\;$ \textbf{0.100} $\quad$ & \textbf{0.103} & $\quad$ \textbf{0.036}  $\quad$ & $\quad$ \textbf{ 0.031} $\quad$ \\ 
Lesvos Island 1st year & $\;$ 0.012 $\quad$ &  $\quad$ 0.013 $\quad$ & $\quad$ 0.002 $\quad$ & $\quad$ 0.0015 $\quad$\\  
Lesvos Island 2nd year & $\;$ 0.007 $\quad$ &  0.008   & $\quad$  0.001 $\quad$ & 0.001 \\
\hline  
\hline
\textbf{Maximum variance} & Plants & Pollinators & Plants & Pollinators  \\
\hline 
Illinois & $\;$ \textbf{0.066} $\quad$ & \textbf{0.069} & $\quad$ \textbf{0.027}  $\quad$ & $\quad$ \textbf{0.028} $\quad$ \\ 
Lesvos Island 1st year & $\;$ 0.015 $\quad$ &  $\quad$ 0.014 $\quad$ & $\quad$ 0.004 $\quad$ & $\quad$ 0.003 $\quad$\\  
Lesvos Island 2nd year & $\;$ 0.011 $\quad$ &  0.011   & $\quad$  0.003 $\quad$ & 0.002 \\
\hline  
\hline
\textbf{Maximum entropy} & Plants & Pollinators & Plants & Pollinators  \\
\hline 
Illinois & $\;$ 0.040 $\quad$ & 0.042 & $\quad$ 0.010  $\quad$ & $\quad$ \textbf{0.015} $\quad$ \\ 
Lesvos Island 1st year & $\;$ 0.009 $\quad$ &  $\quad$ 0.007 $\quad$ & $\quad$ 0.001 $\quad$ & $\quad$ 0.001 $\quad$\\  
Lesvos Island 2nd year & $\;$ 0.006 $\quad$ &  0.007  & $\quad$  0.001 $\quad$ & 0.0016 \\
\hline  
\hline
\end{tabular}
\label{table_ks_twosample}
\end{table}
\end{center}

\begin{center} 
\begin{table}[h]
\caption{Results of the K-L divergence among the synthetic distributions and the empirical distributions of overlap, either among mutualistic or competitive partners. The smaller the KL-divergence, the smaller the difference between the two samples. Further details on the definition and numerical implementation of this measure can be found in the Supplementary Material.}
\centering
\begin{tabular}{c c c c c c} 
\hline
\hline
\textbf{Dataset} / Model &\multicolumn{2}{c}{$D_{KL}$ mutualism} & \multicolumn{2}{c}{ $D_{KL}$ competition}& \\ 
\hline
\hline
\textbf{Illinois} & Plants & Pollinators & Plants & Pollinators &  Average \\
\hline 
Synchronized & 0.035 & 0.038 & 0.002 & 0.014 & 0.022 \\
Minimum competition & 0.095 & 0.101 & 0.026 & 0.033 & 0.064 \\
Maximum variance &	0.042 & 0.056 & 0.010 &	0.019 & 0.032 \\
Maximum entropy & 0.022 & 0.032 & 0.003 & 0.008 & 0.016 \\
\hline  
\hline
\textbf{Lesvos Island 1st year} & Plants & Pollinators & Plants & Pollinators & Average \\
\hline 
Synchronized & 0.0116 & 0.0053 & 0.0014 & 0.0010 & 0.0048 \\
Minimum competition & 0.0050 & 0.0073 & 0.0009 & 0.0004 & 0.0034 \\
Maximum variance & 0.0111 & 0.0047 & 0.0032 & 0.0008 & 0.0049 \\
Maximum entropy & 0.0055 & 0.0051 & 0.0005 & 0.0005 & 0.0029 \\
\hline  
\hline
\textbf{Lesvos Island 2nd year} & Plants & Pollinators & Plants & Pollinators & Average \\
\hline 
Synchronized & 0.0048 & 0.0036 & 0.0008	& 0.0008 & 0.0025 \\
Minimum competition & 0.0018 & 0.0035 & 0.0005 & 0.0005 & 0.0015 \\
Maximum variance & 0.0080 & 0.0036 & 0.0019	& 0.0005 & 0.0035 \\
Maximum entropy & 0.0019 & 0.0035 & 0.0005 & 0.0007 & 0.0016 \\
\hline  
\hline
\end{tabular}
\label{table_kl_divergence}
\end{table}
\end{center} 

In order to differentiate which model better reproduces the empirical phenology in the cases in which, according to the KS test, various models are statistically compatible, we carry out \textit{(b)}, that is, the calculation of the KL-divergence. Indeed, the KL distance measures the amount of information lost when we approximate the empirical distribution by the synthetic one, and therefore it can be used as a statistical criterion for model selection. In Table~\ref{table_kl_divergence} we can observe that the model based on maximum entropy generally provides the best approximation, except for the second year of the Lesvos Island dataset, in which it is slightly outperformed by the model based on minimizing the competition. It is also interesting to observe that the second-best model is not the same among datasets: while the second-best model for the Illinois dataset is the one with synchronized phenology, in the Lesvos Island dataset it is the model which minimizes the competition.

\subsubsection{Structural features}

Secondly, in order to complement the information provided by depicting the phenological coefficients, we calculate the daily networks resulting from considering the phenology. This again produces a sequence of synthetic networks, each corresponding to a given date, as represented in Fig.~\ref{fig_temporal_networks}. Repeating the procedure detailed in section 3.2. now for each synthetic model, we calculated a few basic structural quantities as a function of time: the number of active nodes -size-, the number of active links, and the maximum degree. 

With this information at hand, we can perform a comparison between the product of each synthetic model and the empirical observations, specifically addressing the change in fundamental structural properties over time. To quantify these correlations, we calculated the corresponding Pearson coefficient for different values of time lag between the distributions. By doing so, we are taking into account the possibility that the null model reproduces well the empirical scenario but in a delayed or advanced time. That is, we are relaxing the assumption of perfect matching between the empirical and the synthetic starting dates, focusing instead on the relative position among the starting dates of different species. In table~\ref{table_temporal_pearson} we show the maximum value of the Pearson correlation calculated following this procedure (technical details can be found in the Supplementary Material). 

The study of table~\ref{table_temporal_pearson} reveals that the analysis of the network structure leads to similar results to the aforementioned analysis of the phenological overlap. Indeed, the better fitting models for the Illinois dataset are the ones based on synchronized phenology and maximum entropy, while for the Lesvos Island dataset, the maximal correlation is found for the model which minimizes intra-guild competition and, again, the one maximizing the entropy. 

\begin{center} 
\begin{table}[h]
\caption{Maximum Pearson correlation among the empirical and the null expectation of the structural properties along time.}
\centering
\begin{tabular}{c c c c c } 
\hline
\hline
\textbf{Dataset} / Property & \multicolumn{4}{c}{Maximum Pearson coefficient for each null model}  \\ 
\hline
\hline
\textbf{Illinois} &  Synchronized & Max Variance & Min competition & Max entropy\\
\hline 
Maximum degree & 0.98 & 0.88 & 0.93 & 0.98 \\
Number of links & 0.98 & 0.87 & 0.78 & 0.98 \\
Size & 0.99 & 0.79 & 0.92 & 0.98 \\
Average & 0.99 & 0.85 & 0.88 & 0.98 \\
\hline  
\hline
\textbf{Lesvos Island 1st year} & Synchronized & Max Variance & Min competition & Max entropy \\
\hline 
Maximum degree & 0.78 &	0.76 &	0.91 &	0.91 \\
Number of links	& 0.74 & 0.71 &	0.89 &	0.83 \\
Size & 0.81 & 0.58 &	0.89 &	0.85 \\
Average	& 0.78 & 0.69 &	0.90 &	0.86 \\
\hline  
\hline
\textbf{Lesvos Island 2nd year} & Synchronized & Max Variance & Min competition & Max entropy\\
\hline 
Maximum degree &	0.90 &	0.67 &	0.87 &	0.93 \\
Number of links	& 0.85 &	0.78 &	0.89 &	0.85 \\
Size	& 0.89 &	0.63 &	0.90 &	0.88 \\
Average	 &0.88 &	0.69 &	0.88 &	0.88 \\
\hline  
\hline
\end{tabular}
\label{table_temporal_pearson}
\end{table}
\end{center} 

Overall, these multiple statistical tests reveal at least two general conclusions. On the one hand, we find that each dataset is better described by a distinct mechanistic assumption, i.e. the synchronization of the species' phenologies in the Illinois dataset and the minimization of the competition in the Lesvos Island dataset. This disparity seems to reflect the substantial divergences between the two communities' phenologies, which we described in detail in section 3.1. On the other hand, the temporal networks' features are accurately reproduced by the statistical model that maximizes the entropy associated with the distribution of the middle dates of activity, despite the considerable differences between the two datasets. What this latter finding suggests is that the observed network of interactions together with the periods of each species provide sufficient information to reproduce, fairly closely, the observed temporal patterns of activity. This is particularly interesting given the mentioned differences among datasets, which provide, to a certain extent, a warranty of generality despite the limited data at our disposal. This finding does not imply that, forcedly, the starting dates of activity are set at random, but that the information enclosed in the corresponding network of interactions and the species' periods may be sufficient to reproduce well the main characteristics of the community's phenology.

\section{Conclusions}

Along this paper, we have addressed the challenge of moving beyond the aggregated paradigm by incorporating some of the temporal variability that plant-pollinator communities exhibit throughout the year. In particular, we have concentrated on seasonal ecosystems, assessing the daily change in both inter-guild, mutualistic relationships and intra-guild, competitive interactions. 

The analyses carried out by using both synthetic models and empirical data have revealed, in the first place, that non-trivial information is lost when portraying the real network of interactions by a static representation, that is, neglecting its temporal dimension. Indeed, we have observed that the consequences of introducing the empirical phenology into the network formalism are system-dependent, and hence no general pattern can be expected a priori. Moreover, the process of aggregation not only disregards the richness of temporal variability, but it also tends to overestimate the value of the main fundamental structural features, as had been remarked as well by\cite{sajjad2017effect}. This is especially relevant given that structural features like the degree or the connectivity are consistently used to characterize, respectively, the relative vulnerability of species\cite{dakos2014critical} or the community stability\cite{thebault2010stability}.

In light of the limitations of the aggregated paradigm, and driven by the scarcity of available datasets, we proposed a group of models to produce, given a fixed mutualistic network, a compatible hypothetical phenology. The comparison of these results with the empirical datasets revealed that the soundness of certain mechanisms is, at least in the first-order approach we have adopted here, specific to the particular system under consideration. We have found that the purely statistical assumption of maximizing the entropy associated with the distribution of middle dates performs generally well, as we have tested in two dissimilar empirical examples. Importantly, the remarkable performance of the maximum entropy hypothesis is partly explained by the fact that, actually, the network of interactions is closely dependent upon the starting dates, in the sense that the mutualistic contacts observed corresponded, forcedly, to concurrent species --a condition that, indeed, we impose to our models. Therefore, preserving the network of interactions is a strong constraint, which could justify the general adequacy of this model. 

In perspective, these synthetic models offer a methodological set that might prove useful in different aspects. On the one hand, they can be exploited merely as a group of realistic models to construct synthetic ensembles in the absence of highly-resolved empirical data, in those cases where the main driving forces of the phenology are known. On the other hand, they can also be applied as null models that permit testing a variety of null hypotheses, from the mechanistic forces shaping the phenology to the existence of temporal, structural, or dynamical patterns. At this point, it is worth reminding as well that ours is just a first approach to modeling the temporal variability of networks. In particular, we considered the description level at which species are active or inactive during a certain fraction of the season. However, it could be possible to refine the scale of description to include the weekly or even daily turnover of interactions, a sort of hyper-realistic depiction of the temporal variability of the network that is gaining attention during the recent years\cite{caradonna2017interaction}. This is due, partly, to the technological advances that permit monitoring phenology in great detail, and therefore it is probable that in the coming years, we will find a rising number of this type of studies.

In the bigger picture, the interest in moving towards a more realistic portrayal of interacting systems is not exclusive to ecology, and indeed the study of temporal complex networks has received great attention in recent years\cite{holme2012temporal}. How this change of paradigm will eventually challenge our understanding of natural systems is something we are just now beginning to explore.

\section*{Data availability}
The datasets analyzed in this paper were released by \cite{burkle2013plant} and \cite{kantsa2018disentangling} and are publicly available, respectively, at:

\begin{itemize}
    \item  https://doi.org/10.5061/dryad.rp321\\
    \item  https://figshare.com/articles/dataset/Data\_from\_Disentangling\\
\qquad \_the\_role\_of\_floral\_sensory\_stimuli\_to\_pollination\_networks\_/5663455
\end{itemize}

\section*{Acknowledgments}
The authors would like to acknowledge and thank A. Kantsa for sharing some valuable clarifications on the construction of the Lesvos Island dataset, as well as additional information on the starting and final dates of the pollinator's activity. We acknowledge partial support from the Government of Aragon, Spain, and “ERDF A way of making Europe” through grant E36-23R (FENOL) and from Ministerio de Ciencia e Innovaci\'on, Agencia Espa\~nola de Investigaci\'on (MCIN/ AEI/10.13039/501100011033) Grant No. PID2020-115800GB-I00 to C.G.L. and Y.M.\\
\vspace{1cm}



\bibliographystyle{vancouver}
\bibliography{ref_phen.bib}


\newpage

\begin{center}
\begin{huge}
\textbf{Supplementary Material}\\
\rule{\textwidth}{1pt} 
\\
\end{huge}
\end{center}

\setcounter{section}{0} 
\renewcommand{\thesection}{SM-\arabic{section}} 

\vspace{3mm}

\section{Numerical implementation of synthetic\\models}
\label{Appendix_synthetic_models}

In this section, we explain the different numerical methods used to determine the starting dates of species' activity, on the basis of different hypotheses and under the constraint that mutualistic partners share a minimum phenological overlap.

First, we describe two numerical approaches to solve the problem of maintaining the mutualistic overlap: \textit{(i)} by shifting some given starting dates, \textit{(ii)} by optimizing a certain quantity within the constraints imposed by the non-zero overlap.
Secondly, we explain how we implemented these methods for each particular synthetic model. 

\subsection{Shift of starting dates}
\label{null_model_shift}

Here we introduce a method to shift a given set of initial times of activity, with the condition that a minimal mutualistic overlap is preserved. The periods of activity, that we will denote as $p^P_i$ for a plant $i$ and $p^A_k$ for an animal $k$, are kept unchanged. We solely modify the timing of the beginning of the activity of each species, where we note the initial time by $t^P_{0,i}$ for each plant $i$ and $t^A_{0,k}$ for each animal $k$. 

\begin{figure}[h]
\centering
  \includegraphics[width=\linewidth]{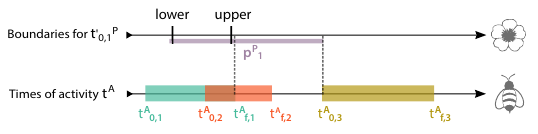}
  \caption{Example of a plant that interacts mutualistically with three pollinator species. We calculate the upper and lower boundaries for the new initial time of the plant $t'^P_{0,1}$ using, respectively, the minimum final time of activity of its pollinators (here $t^A_{f,1}=t^A_{0,1}+p^A_{1}$) and the maximal initial time (here $t^A_{0,3}$) minus the period of the plant ($p^P_1$). As can be seen in the figure, to these boundaries we correspondingly added and substracted one day in order to maintain at least one day of mutualistic overlap.} \label{fig1}
\end{figure}

The program takes as input the aforementioned information together with the starting dates of activity, which, importantly, needs to already fulfill the set of inequalities written in Eqs. 2.9 and 2.10 in the main text. This can be achieved by taking, depending on the case, a trivial solution --e.g. the same starting dates for all species-- or the empirical dataset --if available. Then, the proposal of the novel starting dates is done under the condition that a non-zero amount of temporal overlap between interacting mutualistic partners must be preserved. In detail, we impose, at least, 1 day of overlap, which is the minimal unit of phenology in the studied datasets. To warrant that this occurs, we fix the following boundaries to the new initial times. Let us call $t'^P_{0,i}$ the novel time proposed for plant species $i$:     

\begin{itemize}
\item {The \textbf{upper} bound is given by the neighbor species whose activity ceases earlier. That is, by the minimum of the set of times $\{t^A_{0,k} + p^A_k \}$, where $k$ runs over the indexes of the animal species interacting with plant $i$ (so, if \textbf{B} is the bipartite matrix, those fulfilling that $B_{i,k} = 1$). Let us illustrate this with the example in Fig.~\ref{fig1}. The upper boundary for the initial flowering time of plant number 1 is given by the pollinator that finishes pollinating earlier, which is animal number 1. Therefore, as can be seen in the figure, the upper limit for $\{t^P_{0,1}\}$ coincides with $\{t^A_{f,1}\}$.}
\item {The \textbf{lower} bound is found by subtracting the period of plant \textit{i} to the time at which starts the last species. Thus, by the maximum of the set of times $\{t^A_{0,k}\}$, minus  $p^P_i$. In the example of Fig.~\ref{fig1}, the pollinator that begins its activity the latest is animal number 3, at $\{t^A_{0,3}\}$. To $\{t^A_{0,3}\}$, we need to subtract the period of plant number 1 (in yellow) in order to recover the lower limit.}
\end{itemize}   
 
According to these criteria, the new starting dates of activity are extracted from the following uniform distribution:
\begin{equation}
t'^P_{0,i} \: \subset \:\textit{\Large{U}} \, \Big( \, \textbf{max} \big(\{ t^A_{0,k}\}\big) \, - \: p^P_i  +  1 \, , \,\textbf{min} \big(\{ t^A_{0,k} + p^A_k \}\big) - 1 \, \Big), 
\end{equation} 

where the 1 is respectively subtracted and added to the boundaries in order to ensure at least one day of overlap between mutualistic partners.

Note again that this procedure requires an initial configuration of starting dates, either empirical or hypothetical, that is then randomized. Subsequently, the resulting set of starting dates shows some dependency on the initial configuration. 

As explained in the main text, we exploit this feature to generate synthetic phenologies that are approximately centered. In order to produce a configuration where species' periods are relatively synchronized, we proceed in several steps. First, we set the middle dates by sampling a normal probability distribution. Once the medium dates are determined, the location of the period of each species along the seasons is randomized using an algorithmic procedure that conserves the mutualistic interactions provided by the empirical network. 

In more detail, we programmed a software in Python that operates as follows:

\begin{itemize}
\item We select one of the guilds (for instance, the pollinators) and determine their corresponding starting dates by extracting their middle dates from a normal distribution with standard deviation $\sigma=0.5$ days and mean $\mu=120$ days.
\item We then set the starting dates for the other guild, by using the randomization process explained \LH{above.} In detail, we determine the range -maximum and minimum dates- within which we can extract a random starting date while ensuring that the mutualistic overlap will not be lost. This novel starting date is extracted from a uniform distribution.
\item To finish, we re-randomize again the first guild, using the same procedure as in the previous step, in order to relax the assumption of the normality of middle times of activity.
\end{itemize}

Indeed, we repeat steps \textit{(ii)} and \textit{(iii)} iteratively several times in order to improve the randomization. This eventually yields a non-perfectly synchronized configuration of phenology.

\subsection{Constrained optimizations}
\label{app_constrained}

Here we describe a second procedure to generate a set of starting dates that fulfills the inequalities described in Eqs.~2.9 and 2.10 in the main text. In contrast with the algorithm described in section~\ref{null_model_shift}, this method does not require any other input than the network of mutualistic interactions and the periods, and hence there is no initial condition for the starting dates. Instead, we seek a solution that, on the one hand, verifies the system of coupled inequalities, and on the other hand, optimizes a given quantity, described by an objective function. 

In particular, given the constraints in Eqs.~2.9-2.10, we may use linear programming techniques to find a feasible solution. Since $L>>N$, the problem is overdetermined, yet we know already that it is consistent because there are some trivial solutions that satisfy all the constraints -e.g., all species starting or finishing the same date-. However, we are particularly interested in non-trivial solutions that can lead to realistic or statistically interesting configurations of phenology. This leads to the introduction of the objective function, which we will seek to optimize over the possible set of solutions. In sections~\ref{app_mincomp}-\ref{app_maxent} we specify the numerical implementation of different forms of this objective function, but the general structure of the algorithm is maintained.

In detail, we solve the constrained optimization problem defined by Eqs.~2.9-2.10 through a two-step process. First, we perform multiple local optimizations with different initial seeds, using the function \textit{minimize} from the \textit{SciPy} package in Python\cite{2020SciPy} --specifically the method '\textit{trust-constr}'. Second, we set these local solutions as the starting population for a differential evolution algorithm, which performs a global stochastic search. In particular, we use the \textit{differential\_evolution} function from \textit{SciPy}. This combination of an iterative local search plus a global optimization aims at ensuring a robust finding of the global optimum.

\subsection{Minimization of the competition}
\label{app_mincomp}

In order to find the configuration that, under the mentioned constraints, minimizes the competition, we apply the numerical algorithm described in section~\ref{app_constrained} where the objective function is the total competitive overlap. In detail, at each iteration of the optimization algorithm, we calculate the total sum of unnormalized phenological overlaps among all competitors, including both competition among plants and among pollinators. 

\subsection{Maximization of the variance}
\label{app_maxvar}

We perform a constrained search that maximizes the variance of the middle dates of both guilds by using the algorithm described in section~\ref{app_constrained}. In this case, the objective function corresponds to the variance, and it is calculated at each iteration of the optimization process. 

\subsection{Maximization of the entropy}
\label{app_maxent}

In this model, we find the phenological configuration that maximizes the entropy by using the algorithm explained in section~\ref{app_constrained}. The objective function corresponds here to the Shannon-Gibbs entropy of the middle dates, as defined in the main text, which we explicitly computed using the Python software.

\section{Statistical Tests}
\subsection{Quality of a fit using the Kolmogorov-Smirnov test}
\label{ks_test_one_sample}

In order to assess the quality of the fit and obtain its corresponding p-value, we start by performing a Kolmogorov-Smirnov one-sample test between the fitted function and the empirical distribution, using the \textit{stats.kstest} software from the \textit{SciPy} package\cite{2020SciPy} in Python. Next, in order to determine its p-value we explicitly compare the obtained KS-value with the expected statistics, sampled using Monte Carlo simulations. 

In more detail, we proceed as follows:
\begin{itemize}
\item We calculate the K-S test statistics --using the SciPy software--, that we may call $D_{obs}$, between the empirical distribution and the fitted function, whose parameters are estimated using the maximum likelihood fitting software from the \textit{stats} function in SciPy.
\item Using Monte Carlo methods, we generate a sample of size $N_{sampl}$ composed by synthetic distributions, each sampled from the fitted function and having the same size as the observed empirical distribution. 
\item For each synthetic sample indexed by $i$, we fit a function using the same functional form as in the empirical case and using the maximum likelihood estimator from the \textit{stats} function in SciPy.
\item We then calculate the K-S test statistics between this novel fit and the corresponding synthetic sample, which produces a new distance that we will call $D_{syn,i}$.
\item Once this procedure has been repeated for all the synthetic distributions generated, we can compare the observed K-S statistics $D_{obs}$ with the distribution of $D_{syn,i}$ in the sample.  
\end{itemize}

After carrying out this computation, the p-value is then calculated as:

\begin{equation}
\text{p-value} \, = \, \frac{\text{number of samples with } D_{syn,i} > D_{obs} \, + \, 1}{N_{sampl}},
\end{equation}

which, as normally, quantifies whether the observed KS distance $D_{obs}$ differs significantly from what would be expected if the observed distribution of periods was really generated by the fitted function. Hence, when the p-value is sufficiently large the fit is compatible with this assumption.

\subsection{Kolmogorov-Smirnov two-sample test}
\label{app_ks_twosample}

We programmed a Kolmogorov-Smirnov test on two samples by using the \textit{ks\_2samp} function from the \textit{SciPy} package.
In detail, the K-S test for two samples permits challenging the null hypothesis that two particular samples come from the same statistical distribution. The smaller the KS distance, the smaller the discrepancy between the two samples. In the case where the p-value is relatively small (e.g. p-value $<0.05$) the difference among distributions is significant and hence the null hypothesis is rejected, which in our case implies that the distribution produced by the given synthetic model is incompatible with the empirical observations.

\subsection{Kullback-Leibler divergence}
\label{app_kl}

The Kullback-Leibler divergence quantifies how one probability distributions $Q(x)$ is different from a second, reference distribution called $P(x)$. In particular, the KL divergence from Q to P is defined as:
\begin{equation}
D_{KL} \, = \, \sum_x P(x) \, log \frac{P(x)}{Q(x)},
\end{equation}

and it measures the amount of information lost when approximating $P(x)$ by $Q(x)$. In our case, we calculate this quantity by taking as the reference function $P(x)$ the empirical distribution of phenological overlap, while $Q(x)$ corresponds to the distribution of phenological overlaps produced by the null model under consideration. We implement this calculation in a program in Python that uses the \textit{stats.entropy} software from the \textit{SciPy} package.

\subsection{Pearson correlation with time lag}
\label{app_pearson_time}

In order to calculate the Pearson correlation for different values of time lag, we take the evolution of a given structural property in the synthetic model, and shift it alternatively backwards or ahead in time. For each different value of time lag, we calculate the corresponding Pearson coefficient between the synthetic, shifted array, and the unchanged empirical vector. In particular, we use the \textit{stats.pearsonr} software from the \textit{SciPy} package to compute the Pearson correlation. We repeat this procedure for a wide range of different temporal lags (from zero to a full season), and eventually keep the maximum value of the Pearson coefficient. This provides the time lag at which the correlation between the empirical and the synthetic sequences is maximized, as well as the value of this correlation. 



\end{document}